\documentclass[aps,twocolumn,showpacs,superscriptaddress]{revtex4}
\usepackage{graphicx}
\begin{document}
\newcommand{\beq}{\begin{equation}}
\newcommand{\eeq}{\end{equation}}
\newcommand{\ben}{\begin{eqnarray}}
\newcommand{\een}{\end{eqnarray}}
\newcommand{\bea}{\begin{array}}
\newcommand{\eea}{\end{array}}
\newcommand{\om}{(\omega )}
\newcommand{\bef}{\begin{figure}}
\newcommand{\eef}{\end{figure}}
\newcommand{\leg}[1]{\caption{\protect\rm{\protect\footnotesize{#1}}}}
\newcommand{\ew}[1]{\langle{#1}\rangle}
\newcommand{\be}[1]{\mid\!{#1}\!\mid}
\newcommand{\no}{\nonumber}
\newcommand{\etal}{{\em et~al }}
\newcommand{\geff}{g_{\mbox{\it{\scriptsize{eff}}}}}
\newcommand{\da}[1]{{#1}^\dagger}
\newcommand{\cf}{{\it cf.\/}\ }
\newcommand{\ie}{{\it i.e.\/}\ }


\title{Broken Ergodicity in classically chaotic spin systems}

\author{F.Borgonovi}
\email{f.borgonovi@dmf.unicatt.it}
\affiliation{Dipartimento di
Matematica e Fisica, Universit\`a Cattolica, via Musei 41, 25121
Brescia, Italy}
\affiliation{I.N.F.M., Unit\`a di Brescia and 
I.N.F.N., Sezione di Pavia, Italy}
\author{G.L.Celardo}
\email{celardo@dmf.unicatt.it}
\affiliation{Dipartimento di
Matematica e Fisica, Universit\`a Cattolica, via Musei 41, 25121
Brescia, Italy}
\affiliation{I.N.F.M., Unit\`a di
Milano, Italy}
\author{M.Maianti}
\affiliation{Dipartimento di
Matematica e Fisica, Universit\`a Cattolica, via Musei 41, 25121
Brescia, Italy}
\author{E.Pedersoli}
\affiliation{Dipartimento di
Matematica e Fisica, Universit\`a Cattolica, via Musei 41, 25121
Brescia, Italy}

\date{\today}

\begin{abstract}
A one dimensional classically chaotic spin chain with asymmetric
coupling and two   
different inter-spin interactions, nearest neighbors 
and all-to-all, has been considered.
Depending on the interaction range, dynamical 
properties, as ergodicity and chaoticity are strongly different. 
Indeed, even in presence of chaoticity, the model displays a lack 
of ergodicity only in presence of all to all interaction
and below an energy threshold,
that persists in the thermodynamical limit. 
Energy threshold can be found analytically and results
can be generalized for 
a generic  $XY$ model with asymmetric  coupling.

\end{abstract}
\date{31.10.2003}
\maketitle

\section{Introduction}

Systems with few degrees of freedom
but good chaotic properties, can be  characterized by
standard statistical properties, for instance  diffusive
random-walk like behavior\cite{boris} or appearance of
stationary distribution for single particle occupation numbers
\cite{fla,bi,bigc}.

It is also generally assumed that, 
due to a sufficiently strong interaction, chaos will provide 
the mechanism in order to have ergodicity on the energy surface.
Generally speaking, the degree of chaos will depend on the strength
of the interparticle interaction and on the number of particles:
typically, the larger the number of particles,
the stronger is chaos,  thus leading to more suitable conditions 
for ergodicity and statistical investigation.

This common lore is far from being satisfied by most physical systems,
as explained by R.G. Palmer in his seminal paper\cite{Palmer}
 where, using his words: ``The breakdown of ergodic behavior 
is discussed as a general phenomenon in condensed matter physics''.
In his paper  the breakdown of ergodicity
as due to an ``effective'' confinement of the system, respect to 
the observational timescale has been put forward.
Nonetheless, to the best of our knowledge, there are no single
examples where thresholds for broken ergodicity can be found analytically
and, more important, where the relation with the interaction range has
been remarked explicitly. 

Only in recent years the relation between the dynamical
properties, like chaos or ergodicity, and the kind of interparticle
interaction has been studied \cite{tsallis0}.
Indeed, long-range interacting systems display peculiar
properties from the point of view of statistical mechanics.
It is well known for instance that, for such systems, the canonical
and the microcanonical ensemble give different results, \cite{ruffo},
thus questioning the validity of statistical mechanics.
Moreover, chaos suppression 
in long-range interacting systems
when the number of particles increases
was recently found\cite{ruffo0, tsallis1, tamarit},
and analytically investigated \cite{Firpo},
thus leading to anomalous statistical behavior in the 
thermodynamical limit.

In this Paper, we investigate a  one--dimensional
spin chain model with two different ranges of interaction,
nearest neighbors and  ``infinite'' (all particles are interacting with
all the others)
and we  show
that, even in presence of dynamical chaos\cite{nota1},
a non--ergodic behavior is found, for any finite number of particles.
Such non--ergodic behavior is mainly due to a unconnected  
phase space and the threshold for non-ergodicity can be obtained
analytically. Moreover such non-ergodicity persists in the
thermodynamical limit, actually the ratio
between the disconnected portion of the energy range 
and  the whole one approaches one, as the number of particles
increases.

We have also found that,
for all-to-all interaction, when the number of particles
increases, the system dynamics becomes more regular.
We also give numerical evidence that there is room for chaos
in the non--ergodic region, for any finite number of particles.

For sake of comparison, we also consider the same  model  
with short range interaction (nearest neighbor coupling) and we 
prove  that this kind of disconnection does not  exist.
Of course, this does not imply that the system is ergodic,
as also supported by our numerical simulations that indicate
the presence of regular dynamics in some
region of the energy space.

\section{The Model : chaoticity {\it vs} interaction range} 

Our model is a variant of the one--dimensional Heisenberg model
for $N$--spins. The Hamiltonian is given by:

\begin{equation}
\label{ham}
H= J \sum_{\langle i,j\rangle} (s_i^xs_j^x-s_i^ys_j^y)
\end{equation}

where $\langle i,j\rangle$ stands for nearest neighbor
($\cal N$--interaction),
or infinite range  couplings (all--to--all, $\cal A$--interaction) 
and $J$ is a positive constant.

Hamiltonian (\ref{ham}) gives rise to the standard 
nonlinear equations of motion:

\begin{equation}
\label{eom}
\left\{ \begin{array}{lll}
\dot{s}_i^x &=& -Js_i^z \sum_{\langle j\rangle} s_j^y\\
\\
\dot{s}_i^y  &=& -Js_i^z \sum_{\langle j\rangle} s_j^x\\
\\
\dot{s}_i^z &=& J \sum_{\langle j\rangle} (s_i^ys_j^x+s_i^xs_j^y),
\end{array}\right.
\end{equation}
 
where $\langle j\rangle$ is a shorthand notation 
for  $ j \ne i$ (in the case of 
 $\cal A$--interaction) , and $j = i \pm 1$ when $\cal{N}$
interaction (without periodic boundary conditions) is assumed.

Constants of motion are the energy $H=E$,
and the $N$ squared moduli $|\vec {s}_i|^2$ (which we set
equal to $1$ for simplicity).
For $N=2$, an additional  constant of motion, in involution
with the others, is given by $s_1^z-s_2^z$, 
and the  system becomes exactly integrable.

Let us notice that, in general,
the only free parameters are the total energy $E$, the interaction
strength $J$ and the number of particles $N$.
Moreover, for any finite 
number of particles $N$, the  energy is bounded $|E| \leq E_{max} (J,N)$.
One can also give a rough estimate for such border as
$E_{max} \sim JN^2$ for $\cal A$-interaction and $E_{max} \sim JN$
for $\cal N$-interaction, even if
the rigorous bound will be given below.

In order to make a fruitful  comparison between
$\cal N$ and $\cal A$ interaction
and according to a general prescription \cite{ruffo1} we rescale,
for $\cal A$ interaction, the strength $J$ to the number of particles $N$,
setting $J=2I/N$.
In this way
the  energy of  $\cal A$ systems scales with the number
of particles in the same way as
for  $\cal N$ systems.

For $N>2$,  and sufficiently strong interaction 
(both $\cal N$ and $\cal A$ interaction)
the system is chaotic in a large energy range 
$|E| < |E_{ch}(j,N)| < E_{max}$, as indicated by 
 a maximal positive Lyapunov 
exponent.
From now on, all results will be restricted to the 
chaotic energy region only.

\begin{figure}
\includegraphics[scale=0.46]{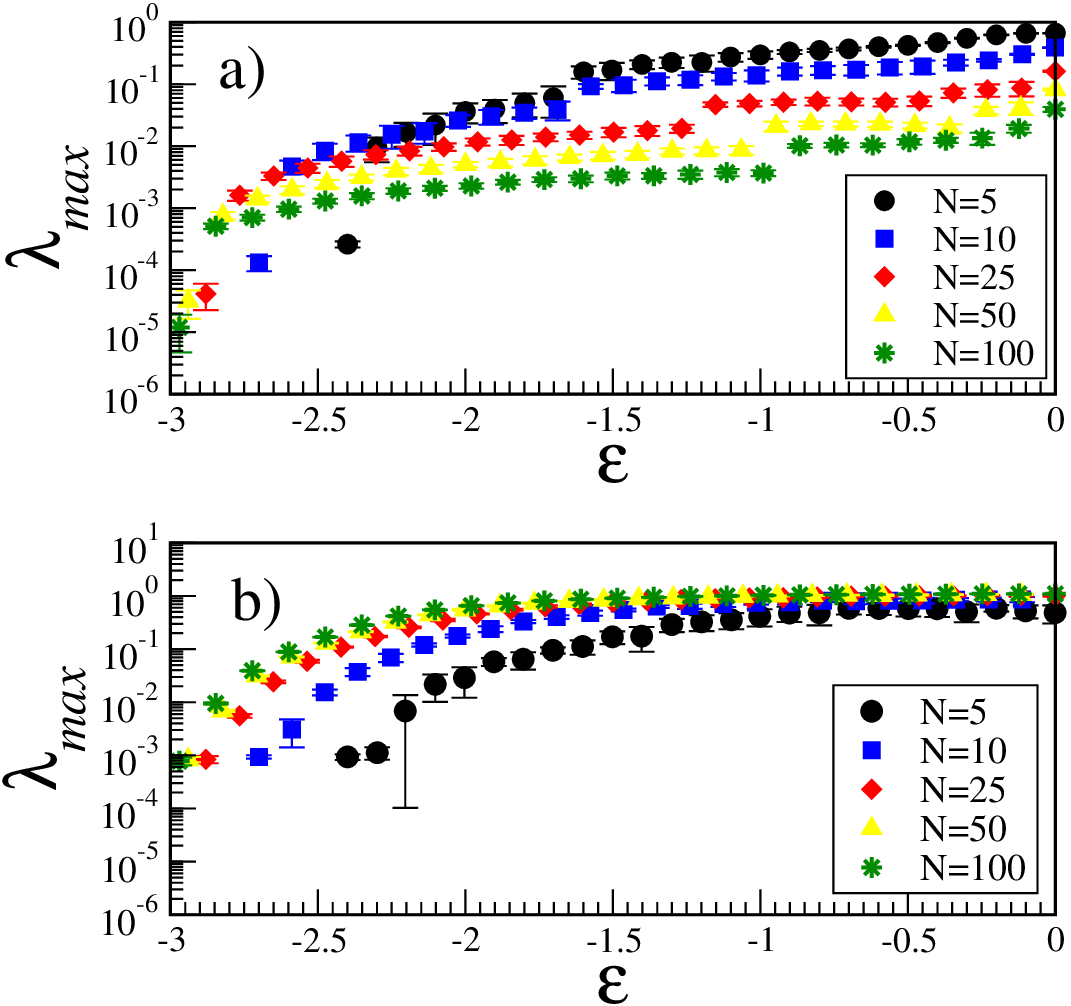}
\caption{
a) Maximal Lyapunov exponent, $\lambda_{max}$, for
different number of particles $N=5$ (black) 
, $N=10$ (blue), $N=25$ (red), $N=50$ (yellow) and $N=100$ (green), 
as a function of the energy per particle $\epsilon=E/N$,
for $\cal A$--interaction $a)$, and for 
$\cal N$--interaction $b)$. Error bars have been obtained 
as the  standard deviation from a set of $20$ initial trajectories on
the same energy surface.
Trajectories have been integrated for a time of $10^4$.
In $a)$ the interaction constant has been chosen as $J=2I/N$,
with $I=3$,
while in $b)$ $J=I=3$.
}
\label{liap}
\end{figure}

\begin{figure}
\includegraphics[scale=0.40]{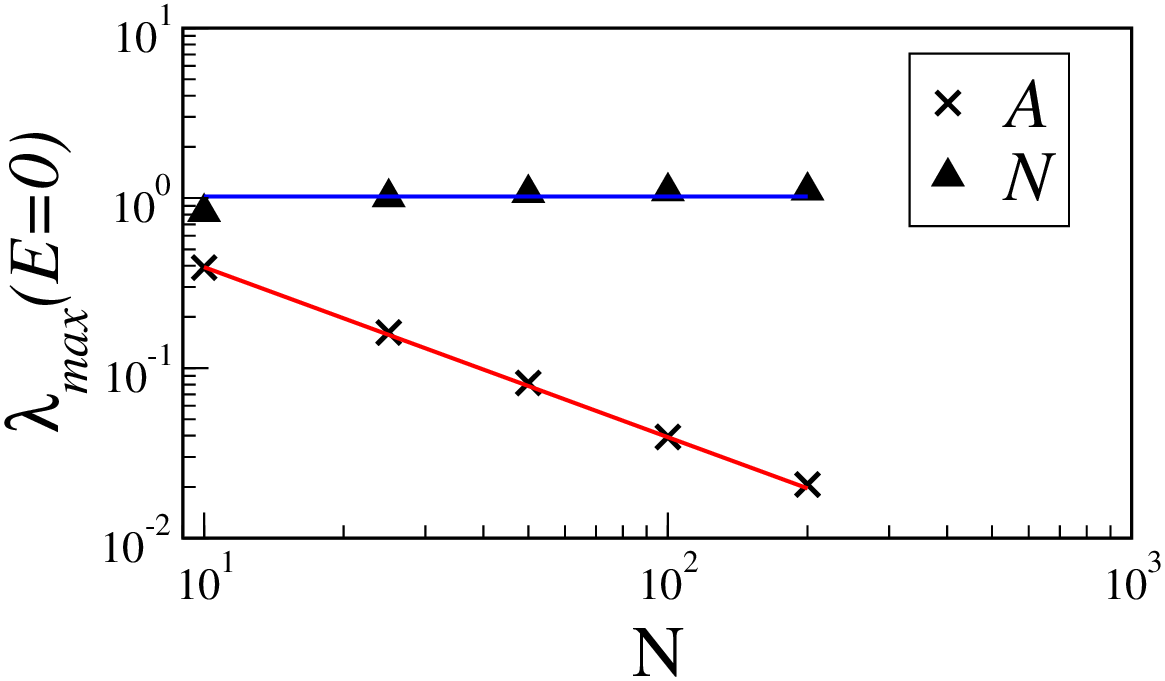}
\caption{Maximal Lyapunov exponent as a function of the number of
 particles, for the same parameters of Fig.(\ref{liap}) and $E=0$. 
Triangles refer to $\cal A$ interaction, while  crosses
are  for ${\cal N}$ interaction. Also shown the best fit lines
$\lambda_N=1.1\pm 0.2$(blue) and $\lambda_A = (4.1\pm 0.1)/N$ (red).}  
\label{tre}
\end{figure}

Chaotic properties of the model has been analyzed within the frame 
of Standard Lyapunov analysis\cite{benettin,lichtenberg}.
In Fig.(\ref{liap}) we show the maximal Lyapunov 
exponent  as a function
of the energy per particle $\epsilon = E/N$ 
for different number of particles
and $\cal A$ (upper) and $\cal N$ (lower) interaction. 
As one can see, while the maximal Lyapunov exponent 
$\lambda_{max}$ depends on the
energy per particle in some peculiar way, there is a large energy
region where  it is appreciably different from zero for both
interactions.
From the same picture it is clear that while increasing 
the number of particles
in the case of
$\cal N$ interaction,
 does not produces an
appreciable variation of  $\lambda_{max}$, 
the same variation in 
$N$, for 
$\cal A$ interaction, effectively decreases the value 
of $\lambda_{max}$.
Even if it is difficult to give an exact scaling 
relation of the maximal Lyapunov exponent with all
the system parameters ($E, N, J$) the approximate relation 
$\lambda_{max}(E=0, I=3) \sim 1/N^\alpha$ holds, where $\alpha=0$ 
for $\cal{N}$ interaction and $\alpha= 1$
for $\cal{A}$ interaction, see  Fig.~(\ref{tre}).

The numerical results and the corresponding approximate relations
are shown in 
Fig.(\ref{tre}) and indicate 
the absence  of a  chaotic dynamics  for
all-to-all interacting systems with a large number of particles.

The general trend where the maximal Lyapunov exponent decreases
with the number of particles can be understood from the
fact that   in the limit $N\to\infty$ and 
for $\cal A$ interaction the model
becomes close to an integrable one.

Indeed the Hamiltonian per particle  can be written as: 

\begin{equation}
\label{hamr}
\frac{H}{N} = I h_0 + \frac{I}{N} h_1 
\end{equation}

where 
$ h_0= m_x^2 - m_y^2$ and 
$h_1 = \frac{1}{N} \sum_{i=1}^{N}\left[ (s_i^y)^2 - (s_i^x)^2 \right]$
where we have defined the average magnetization 
$m_k = \frac{1}{N} \sum_{i=1}^N s_i^k,$ with $k=x,y$.

While both $h_0$ and $h_1$ remain on order 1 in the limit 
of large $N$, and fixed interaction strength $I$,
the constant in front of $h_1$ goes to zero and   
the  $h_0$ term dominates. 
A close inspection\cite{prep}, and our numerical results
indicate that $h_0$ represents an integrable
model for any choice of parameters.

\section{Ergodicity : numerical results}

While in two degrees of freedom systems the absence
of ergodic motion in presence of chaos is somehow
obvious (due to the presence of invariant tori),
for  many degrees of freedom systems the same 
occurrence  is far from being
trivial. 
In spite of that, 
we will now show that, for $\cal A$ interaction, an
energy  threshold exists below which one cannot have ergodic motion.

Let us first evaluate ensemble and time averages for the 
$y$ magnetization and from them, respectively, the
probability distributions $P^p(m_y)$ and
$P^t(m_y)$.

\begin{figure}
\includegraphics[scale=0.36]{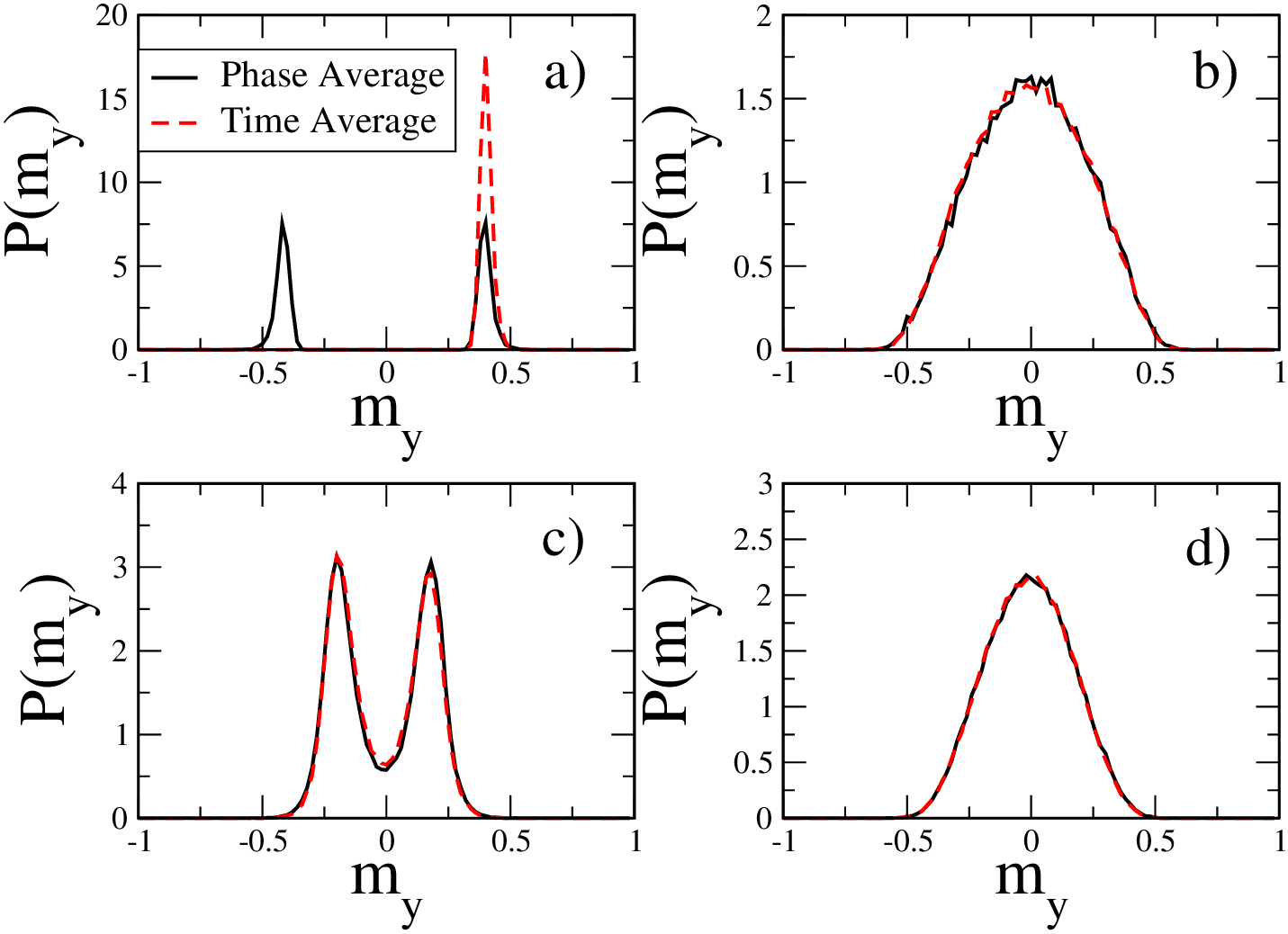}
\caption{Distribution of mean magnetization along the $y$-axis 
for $\cal A$ interaction (left column) 
and $\cal N$ interaction (right column).
As  black lines we show the phase distributions, $P^p(m_y)$,
while red ones
stand for time distributions, $P^t(m_y)$.
Phase distributions have been obtained taking 
$M=10^5$ different initial conditions upon
the energy surface $E=-5$,  $a)$ and $b)$
and $E=-1$,  $c)$ and $d)$,
while time averages have been obtained by computing one single
trajectory in time on the same energy surface for a time
$T=10^5$ and taking values at time steps  $\Delta t=1$.
Results have been checked  on increasing the number of
initial points upon the energy surface for the phase average, and
on increasing the integration time $T$ for the time average.
Here is $N=10$, and $I=3$. 
Maximal Lyapunov exponents are positive for all the 
cases shown in this figure, as can be deduced from 
Fig.(\ref{liap}).}
\label{due}
\end{figure}

Phase and time distributions of the mean magnetization
are shown in Fig.(\ref{due}) for the $\cal N$-interaction (right
column) and $\cal A$-interaction (left column)  
with the same number of particles.
All cases are characterized by a chaotic dynamics,
as given by a maximal positive Lyapunov exponent, see Fig.~(\ref{liap}).

As one can see, while for nearest neighbor interaction
there is a good correspondence between the two averages
(this of course does not mean ergodicity), in the case
of all-to-all interaction there are strong deviations,
in the lower energy case, see Fig.(\ref{due}$a$).
More precisely one trajectory with an initial $m_y > 0 (< 0) $ 
cannot reach  a region with   $m_y < 0 (>0) $, below some
energy threshold.
Another difference is related to the presence 
of a two-peaks distribution: this indicates the presence
of a phase transition and it will be the subject
of a separate paper.

Strictly speaking, from these numerical results one can only infer
that the time of transition from one peak to the other one is much 
larger than the simulation time. Nonetheless, in the next section
we will show that such transition time does not exist.
The origin of this deviation is that for those energy values
less than the energy threshold
the phase space is metrically decomposable\cite{khinchin} thus 
making the $\cal{A}$ system rigorously  non-ergodic.

\section{Non--ergodicity border : analytical results}

The existence of a threshold for the non ergodicity  
in case of ${\cal A}$ interaction
can be proved by the following argument. 
Let us assume that initially 
the  spins are oriented in such a way that 
 $m_y(0) >0$ with the  energy $E$.
It is clear that if we are able to prove that 
when  $E< E_{ne}$  one cannot have  
the solution $m_y=0$
on the energy surface $H=E$,
 it means that the trajectory should keep, for any
time $m_y(t) >0$.
Since reversal of $y$ single spin components results in 
change of $y$ magnetization but not of the energy,
the same argument can be applied to $m_y(0)<0$ as well. In other words
the space is metrically decomposable, since one
trajectory with $m_y>0$ cannot reach the region where $m_y<0$
(and viceversa), although both regions belong to the same energy 
surface.

In order to find the explicit value of $E_{ne}$ one should 
consider the problem of minimizing the value of
the energy $E$ under the constraint 
$m_y = (1/N) \sum_{i=1}^N s_i^y = 0$.
In principle this task can be faced by means of the Lagrange multipliers.
Nevertheless from a very simple argument we can 
estimate  the critical
energy $E_{ne}$.

Let us first derive the minimal energy, $E_{min}$, and then
the  non-ergodic
energy, $E_{ne}$, for both  $\cal N $ and  $\cal A$ systems.

\subsection{Minimal energy}
 
We can find the minimal energy by  minimizing the
 Hamiltonian (\ref{ham})  under the constraint:
\begin{equation}
 (s_i^x)^2+(s_i^y)^2+(s_i^z)^2=1, \hspace{0.2cm} \qquad i=1,...,N
\label{Constr1}
\end{equation}

We can take conditions (\ref{Constr1}) into account,
switching to spherical coordinates:

\begin{equation}
\left\{ \begin{array}{lll}
s_i^x &=& \sin(\theta_i) \cos(\phi_i)\\
s_i^y &=& \sin(\theta_i) \sin(\phi_i)\\
s_i^z &=& \cos(\theta_i).
\end{array}\right.
\end{equation}

We can now rewrite Hamiltonian (\ref{ham}) as:
\begin{equation}
 H=J \sum_{\langle i,j\rangle} \sin(\theta_i)\sin(\theta_j) \cos(\phi_i+\phi_j)
\label{Add}
\end{equation}

where 

\begin{equation}
J = \left\{ \begin{array}{ccc}
(2I/N) &{\rm for}& {\cal A-} {\rm interaction} \\
I &{\rm for}& {\cal N-} {\rm interaction} 
\end{array}\right.
\end{equation}

One  absolute minimum value is obtained, for instance,
setting:
\begin{equation}
\left\{ \begin{array}{lll}
 \sin(\theta_i) &=& 1 \\
\cos ( \phi_i+\phi_{j} )  &=& -1
\end{array}\right.
\label{phi}
\end{equation}

where $j=i+1$ for $\cal N$ system and $j \ne i$ for $\cal A$ system.

While the first  equation  of (\ref{phi}) can 
be easily satisfied by both $\cal N$ and $\cal A$ system,
assuming
\begin{equation}
\theta=\frac{\pi}{2}
\label{phi1}
\end{equation}
some care should be taken in order to fulfill the second one.

For $\cal N$ system, 
we can put for instance: 

\begin{equation}
\left\{ \begin{array}{lll}
 \phi_i  &=& 0 \hspace{0.2cm} \quad {\rm for} \quad i \quad {\rm even}  \\
 \phi_i  &=& \pi \hspace{0.2cm} \quad {\rm for} \quad i \quad {\rm odd}  
\label{thetan}
\end{array}\right.
\end{equation}

On the contrary, for
$\cal A$ system, we set  

\begin{equation}
\phi_i=\pi/2 \qquad  {\rm for} \quad i=1,...,N.
\label{thetaa}
\end{equation}

Taking into account that
there are $N-1$ couplings  in  the $\cal N$ system 
and  $N(N-1)/2$ in the $\cal A$ system,
the minimal energy is given by (for both interactions):

\begin{equation}
\label{min}
 E_{min}=-I(N-1).
\end{equation}

\subsection{Non--ergodic threshold} 

We define the non--ergodic threshold $E_{ne}$ in such a way that for
any energy $E<E_{ne}$ the system is metrically decomposable. 
In order to find this non-ergodic (or disconnection) energy we 
have to find the
minimum of Hamiltonian (\ref{ham}) under the additional constraint
\begin{equation}
 m_y=0.
\label{Constr2}
\end{equation}

For $\cal A$ system Hamiltonian (\ref{ham}) can be written as,

\begin{equation}
H=-\frac{I}{N} \sum (s_i^x)^2 + \frac{I}{N} \sum (s_i^y)^2 +IN m_x^2 - IN m_y^2
\label{Hamm}
\end{equation}

Let us now search for the minimum of (\ref{Hamm}) under
the constraints (\ref{Constr1}) and (\ref{Constr2}).
In (\ref{Hamm}) the only term  that can be negative
under the given constraints is the first one, therefore,
$$
E_{ne}\ge -I
$$
If it is possible to minimize this term putting at the same time
all the other  
to zero, we get  the non-ergodic energy.

When $N$ is even we can put half  spins 
with  $s_i^x=+1$, and  half such that $s_i^x=-1$.
In this way  $m_x=0$, $s_i^y=0$ and $m_y=0$, so that:
$$
E_{ne}= -I, \hspace{0.2cm} {\rm for}  \hspace{0.05cm} N \hspace{0.05cm}
{\rm even} 
$$

When $N$ is odd we cannot minimize the first term in (\ref{Hamm}), 
and at the same time let the other terms to be zero, so that:
$E_{ne}>-I$ , for $N$ odd.

Anyway it is easy to give an upper bound for $E_{ne}Â$ 
which is sufficient for our scope.
Indeed we can arrange  $N-1$ spins as in the previous case ($N$ even), 
and for the last spin we assume $s_N^z= \pm 1$.
In this way we have $m_x=0$, $s_i^y=0$ and $m_y=0$, so that:
$$
-I < E_{ne} \le -I+I/N, \hspace{0.2cm} {\rm for} \hspace{0.05cm} N 
\hspace{0.05cm} {\rm odd}
$$

For $\cal N$ system, the non-ergodic energy,
if defined as the minimum energy for which Eq.(\ref{Constr2}) holds,
coincides with the minimal energy.
Indeed the same values of $\phi_i$ (\ref{thetan}) and
$\theta_i$ (\ref{phi1}) which
minimize the Hamiltonian, satisfy condition (\ref{Constr2}) too, so that:
$$
E_{ne}=E_{min}= -I(N-1)
$$
This of course does not exclude the possibility to have some other constraint
that produces a disconnected phase space (even if such conclusion is not 
supported by our numerical simulations).

In Tables \ref{tab1} and \ref{tab2} we summarize the minimal and non-ergodic energies
found.

\begin{table}
\begin{tabular}{||c|c|c||}  \hline \hline
$\cal A$   & $E_{min}$ &  $E_{ne}$ \\ \hline
 $N_{even}:$ & $-I(N-1)$ & $-I$ \\ \hline
 $N_{odd}:$ & $-I(N-1)$ & $( -I, -I+I/N ]$ \\ \hline
\end{tabular}
\caption{Minimal and non-ergodic energy for $\cal A$ system.
Note that in the case of odd number of particles we give a lower
and an upper bound for the non-ergodic energy.}

\label{tab1}
\end{table}

\begin{table}
\begin{tabular}{||c|c|c||}  \hline \hline
$\cal N$   & $E_{min}$ &  $E_{ne}$ \\ \hline
 & $-I(N-1)$ & $-I(N-1)$ \\ \hline

\end{tabular}

\caption{Minimal and non-ergodic energy for $\cal N$ system.}
\label{tab2}
\end{table}

As an interesting consequence of our  results we note that 
the ratio:
\begin{equation}
r=\frac{|E_{ne}-E_{min}|}{|E_{min}|}
\label{rr}
\end{equation}
between the disconnected portion of the energy range and the total
energy range, in the limit of a large number of particles
at fixed interaction strength,
goes to one for $\cal A$ interaction, and is 
exactly zero for $\cal N$ interaction,
thus showing that in this limit the energy range is completely disconnected 
for $\cal A$ interaction only (note that in the positive
region of the energy range the same argument 
used to show the existence of 
a non-ergodic energy can be applied considering 
$m_x$ instead of $m_y$).

As one can see, for ${\cal A}$ interaction,
 as the number of particles increases 
the region of non--ergodicity does not decrease 
thus indicating that caution should be taken when 
ergodicity is tacitly assumed due to small interactions between 
many particles.

The presence of the non ergodic region is
not related with the chaoticity of the system.
Indeed, since  the Lyapunov exponent increases linearly 
with the interaction strength, $I$,
for any  $N$ we can find an $I_c$ value such that 
there is well developed chaos 
for $E<E_{ne}$ and $I>I_c$,
so that the system is chaotic and non-ergodic,
below $E_{ne}$.

\section{Generalization}

Let us finally mention that the previous 
results can be easily generalized.

Let us consider a generic $XY$ Model:
\begin{equation}
\label{hamxy}
H= -J \sum_{\langle i,j\rangle} (s_i^ys_j^y + \eta s_i^xs_j^x)
\end{equation}
where $|\eta| \le 1$ and $J>0$.
Following the same arguments given above, 
we can show the existence of a non-ergodic energy and 
roughly estimate their dependence 
on system parameters.

We will mainly focus on the  ratio
between the disconnected portion of the energy range and the total
energy range, $r$,  Eq.~(\ref{rr}),  in the limit of a
large number of particles.

For $\cal A$ interaction, we can rewrite the Hamiltonian~(\ref{hamxy}) as:
\begin{equation}
H=\frac{J}{2}  \sum_i \left[ \eta (s_i^x)^2 + (s_i^y)^2 \right] 
-\frac{J}{2} N^2 \left( \eta m_x^2 +  m_y^2\right)
\end{equation}

From this equation we can estimate, for large   $N$
$$
E_{min} \sim -\frac{J}{2}N^2,
$$
As for the non-ergodic energy, following the same 
procedure described in the previous section, we have:

\begin{equation}
E_{ne} \simeq \left\{ \begin{array}{lll}
 -\frac{J}{2}N^2 \eta  &{\rm for}&  \eta > 0 \\
0 &{\rm for}&  \eta =  0 \\
\frac{J}{2}N \eta &{\rm for}&  \eta <  0 \\
\end{array}\right.
\label{ner1}
\end{equation}

so that

\begin{equation}
r  \simeq \left\{ \begin{array}{lll}
 1 - \eta  &{\rm for}&  \eta > 0 \\
1  &{\rm for}&  \eta \leq   0 \\
\end{array}\right.
\label{nrr1}
\end{equation}

As one can see $r=0$ only for $\eta=1$, 
while $r\ne 0$  for asymmetric coupling.

For the $\cal N$ case, we have that 
$$
E_{min} \sim -J(N-1),
$$
while 
$$
-J(N-1) \le E_{ne} \le -J(N-3)$$ 
as one can see 
setting, for instance, 

\begin{equation}
s_i^x = \left\{ \begin{array}{lll}
 1  &{\rm for}& i \ {\rm odd}  \\
-1  &{\rm for}& i \ {\rm even}
\end{array}\right.
\label{sxx1}
\end{equation}

Note that for $N\gg 1$,  
$E_{min} \simeq E_{ne}$
so that $r=0$ for any $\eta$.

Thus, we can conclude that broken ergodicity is generic
for asymmetric coupling and all-to-all interaction.

\vspace{0.5cm}

\section{Conclusions}

In conclusion we have studied a spin chain model from the dynamical
point of view for two different  ranges of inter-spin interaction
(nearest neighbor and all-to-all).
Both models are mainly chaotic, in the sense that the dynamics is characterized
by a positive maximal Lyapunov exponent
in a wide energy region. Nevertheless, in the case
of an infinite range of interaction, the motion 
does not explore the whole energy surface.
This can be understood on the basis of a special topology of the phase space.
An analytical estimate gives an energy threshold below which 
the system is rigorously non--ergodic:
the existence of the  threshold has been also
confirmed by numerical simulations.
Moreover, for any interaction strength, 
on increasing the number of particles the range of energy
values characterized by such non ergodic motion spans a finite
portion of the whole energy range, 
if the interaction has an infinite range.
This could be particularly relevant in the 
study of multidimensional systems, where usually ergodicity
is tacitly assumed for a large number of weak--interacting
particles. 
We have also shown that this phenomenon
is generic for $XY$ model with asymmetric
coupling.

Further studies about the consequences of broken ergodicity
with respect to phase transitions, and the comparison with 
the correspondent quantum system are under current 
investigations\cite{prep}.

\section{Acknowledgments}

We acknowledge useful discussions with 
J.~Barr\`e, L.~Benet, R.~Bonifacio, F.~Izrailev, F.~Levyraz, E.~Locatelli, 
S.~Ruffo, T.~Seligman and R.~B.~Trasarti.



\begin{thebibliography}{}

\bibitem{boris}  B.V.Chirikov, Phys. Rep., {\bf{52}}, 263, (1979).

\bibitem{fla} V. V. Flambaum and F. M. Izrailev, Phys. Rev. E {\bf 56}, 
5144, (1997) 

\bibitem{bi} F. Borgonovi and F. M. Izrailev, Phys. Rev. E {\bf 62}, 
6475, (2000)

\bibitem{bigc} F.Borgonovi, I.Guarneri, F.M.Izrailev and G.Casati, 
Phys. Lett. A {\bf  247} , 140 (1998)

\bibitem{Palmer} R.G. Palmer, 
Adv. in Phys.{\bf 31}, 669  (1982)

\bibitem{tsallis0} A. Campa et al, Phys. Lett. A {\bf 286}, 251, (2001)

\bibitem{ruffo} J. Barr\'e, D. Mukamel, S. Ruffo,  Phys. Rev. Lett. {\bf 87}, 3, (2001)

\bibitem{ruffo0} V. Latora, A. Rapisarda, and S. Ruffo, Phys. Rev. Lett. 
{\bf 80},  692, (1998)


\bibitem{tsallis1} C. Anteneodo, C. Tsallis, Phys. Rev. Lett. {\bf 80}, 24, (1998)
 
\bibitem{tamarit} C. Tamarit, C. Anteneodo, Phys. Rev. Lett. {\bf 84}, 208, (2000)

\bibitem{Firpo} M. C. Firpo, Phys. Rev. E {\bf 57}, 6599, (1998);
M. C. Firpo and  S. Ruffo, Journal of Physics A, 34, L511, (2001);
C. Anteneodo, R.N.P. Maia and R. Vallejos,
Phys. Rev. E {\bf 68}, 036120, (2003).  

\bibitem{nota1} According to the common lore we here call chaotic 
system the ones characterized in some parameters 
range by a positive maximal Lyapunov exponent.


\bibitem{ruffo1}J. Barr\'e, T. Dauxois, S. Ruffo,  Physica A {\bf 295}, 254 (2001)

\bibitem{benettin} G.Benettin, Physica {\bf 13D},  (1984) 211.

\bibitem{lichtenberg} A.J.Lichtenberg and M.A.Lieberman, {\it Regular and Stochastic Motion},
Applied Math. Series 38 Spinger-Verlag (1983)

\bibitem{prep} in preparation.

\bibitem{khinchin} A.I.Khinchin {\it  Mathematical Foundations of
Statistical Mechanics}, Dover Publications, New York (1949)

\bibitem{bcic02}F. Borgonovi, G. Celardo, F. M. Izrailev, G. Casati, Phys. Rev. Lett. {\bf 88}, 5, (2002) 

\end{thebibliography}
\end{document}